# In-plane gate induced transition asymmetry of spin-resolved Landau levels in InAs-based quantum wells

Olivio Chiatti[1], Johannes Boy[1], Christian Heyn[2], Wolfgang Hansen[2] and Saskia F. Fischer[1,3,*]

[1] Novel Materials Group, Institut für Physik, Humboldt-Universität zu Berlin, 10099 Berlin, Germany

[2] Institut für Nanostruktur- und Festkörperphysik, Universität Hamburg, 20148 Hamburg, Germany

[3] Center for the Science of Materials Berlin, Humboldt-Universität zu Berlin, 12489 Berlin, Germany

* Corresponding author

KEYWORDS: Quantum-Hall effect, edge channels, InAs quantum well, magnetotransport, in-plane gates

The cross-over from quasi-two- to quasi-one-dimensional electron transport subject to transverse electric fields and perpendicular magnetic fields are studied in the diffusive to quasi-ballistic and zero-field to quantum Hall regime. In-plane gates and Hall-bars have been fabricated from an InGaAs/InAlAs/InAs quantum well hosting a 2DEG with carrier density of about $6.8 \times 10^{11}\ \mathrm{cm}^{-2}$, mobility of $1.8 \times 10^{5}\ \mathrm{cm}^{2}/\mathrm{Vs}$ and an effective mass of $0.042\, m_e$ after illumination. Magnetotransport measurements at temperatures down to 50 mK and fields up to 12 T yield a high effective Landé-factor of $|g^{*}| = 16$, enabling the resolution of spin-split subbands at magnetic fields of 2.5 T. In the quantum Hall regime, electrostatic control of an effective constriction width enables steering of the reflection and transmission of edge channels, allowing a separation of fully spin-polarized edge channels at filling factors $\nu = 1$ und $\nu = 2$. A change in the orientation of a transverse in-plane electric field in the constriction shifts the transition between Zeeman-split quantum Hall plateaus by $\Delta B \approx 0.1\ \mathrm{T}$ and is consistent with an effective magnetic field of $B_{\mathrm{eff}} \approx 0.13\ \mathrm{T}$ by spin-dependent backscattering, indicating a change in the spin-split density of states.

# I. Introduction

Two-dimensional electron gases (2DEGs) may exhibit topological behavior leading to the observation of the fractional and integral quantum Hall (QH) effects in perpendicular high-magnetic fields at low temperatures [1-3]. The formation of Landau levels (LLs) results in quantum Hall edge channels (QHECs) as chiral 1D states and the inter-edge channel interactions raise continuing research interest as Tomonaga-Luttinger liquids [4] or many-body effects as quantized charge fractionalization [5]. In general, for such studies QHECs may be transmitted or reflected at potential barriers created by top- and split-gates [3, 6-8]. Recent experiments on counterflow edge transport in InAs quantum wells suggest that even in the integer QH regime the microscopic structure of the edge states can differ from that of macroscopic transport experiments and require careful consideration [9]. To date, transversal in-plane electric fields are rarely employed to control the formation of QHEC or the currents in the transition regimes between LLs. However, for InAs-based heterostructures spin-orbit coupling may come into play [10-13] and in-plane fields may become a useful tool to act on spin-dependent transport properties.

Here, we investigate transport from zero- up to high-magnetic fields of the QH regime with only one spin-polarized QHEC (filling factor $\nu = 1$). We study the cross-over from a diffusive electron transport in a wide Hall-bar to diffusive but few-channel quasi-one-dimensional transport in a Hall-bar with a micro-constriction. This constriction can be electrostatically depleted to pinch-off by *symmetric* voltages applied to in-plane gates, which allows us to control the transmission and reflection of QHEC and their interaction in InAs-based heterostructures. In-plane electric fields, which are transversal with respect to the current or edge currents along the constriction, are applied by *asymmetric* in-plane gate voltages.

## II. Experiment

Hall-bars were fabricated by micro-laser photolithography and wet-chemical etching in a shallow inverted $In_{0.75}Al_{0.25}As/In_{0.75}Ga_{0.25}As$ quantum well (InGaAs/InAlAs QW) with an inserted InAs channel. The wafer was grown by molecular-beam epitaxy and consists of an $In_{0.75}Al_{0.25}As/In_{0.75}Ga_{0.25}As$ QW, with a strained, 4 nm thick InAs layer at about 53 nm below the surface [14]. In the following we shorten $In_{0.75}Al_{0.25}As$ and $In_{0.75}Ga_{0.25}As$ to InGaAs and InAlAs, respectively. A sketch of the layer sequence is shown in Fig. 1, together with the calculated conduction band edge profile and the probability density of the two lowest-energy states in the QW. The 2DEG is localized in the narrow InAs QW, with penetration into the InGaAs QW. The characterization of the wafer at $T_{\mathrm{bath}} = 0.25$ K in the dark yields an electron density of $n_{\mathrm{s}} = 4.1 \times 10^{11}\ \mathrm{cm}^{-2}$ and a mobility of $\mu = 1.2 \times 10^{5}\ \mathrm{cm}^{2}/\mathrm{Vs}$. After continuous exposure with an infrared light emitting diode, with wavelength 880 nm, for 30 s the density increased to $n_{\mathrm{s}} = 6.8 \times 10^{11}\ \mathrm{cm}^{-2}$ and the mobility increased to $\mu = 1.8 \times 10^{5}\ \mathrm{cm}^{2}/\mathrm{Vs}$. The constrictions were fabricated in a second step with high-resolution micro-laser photolithography and wet-chemical etching of two trenches in a V-like shape, electrically isolated from the 2DEG, as shown in Fig. 3(a). The width and length of the resulting gap are both approximately 4 µm. The contacts to the 2DEG and the in-plane gates were made by sputter-deposition of an approximately 5 nm layer of titanium beneath an approximately 50 nm layer of gold, without annealing. The samples were mounted on chip-carriers and contacted by wedge-bonding with aluminum wire. The surfaces of the chip-carriers on which the samples are mounted are metallic and can be used as back-gates.

Electric transport measurements were carried out in an Oxford Instruments *Triton* dilution refrigerator, and in an Oxford Instruments *HelioxVL* $^{3}$He-cryostat. The chip carriers were mounted perpendicular to the magnetic field, before inserting the sample holder into the cryostats for cooldown. Stanford Research Systems SR830 or Signal Recovery 7265 lock-in amplifiers were used to measure the longitudinal and transversal voltages simultaneously. The gate voltages were generated by high-precision source-meter units by Keithley, followed by low pass filters with a cut-off frequency of 1 Hz to prevent noise from high-frequency signals.

The band edge profile and the probability density of electron wave functions were calculated by using the self-consistent 1D Poisson-Schrödinger solver by G. Snider [15]. The band parameters for the calculation were obtained from the software provided with [16], the Schottky barrier heights are taken from [17]. The estimates of the effect of strain, under the assumption that InAs layer has the same lattice parameter as InGaAs, are based on data from [18], using a linear interpolation of valence- and conduction band edges between InAs and GaAs well material on GaAs substrate. The parameters used for the calculation are summarized in the Supplementary Material.

## A. Band-profile calculation

The results of band-profile calculations of the heterostructure, with the InGaAs quantum well (QW) and the asymmetrically inserted InAs channel as depicted in Fig. 1(a), are discussed in the following. The band profiles and the probability density of the electronic wavefunctions show that a two-dimensional electron gas (2DEG) localized in the quantum well (QW) is formed. While the wavefunction is centered in the narrow 4 nm-InAs channel, it also penetrates significantly into the InGaAs barrier (see Fig. 1(b)), therefore the transport properties of the electrons are not determined exclusively by the InAs channel. The calculations by the 1D Poisson-Schrödinger solver with *unstrained* InAs show that 48% of the probability density is in the InAs channel, 50% is in the InGaAs quantum well and 1.4% is in the InAlAs spacer. The calculations with *strained* InAs show that the wavefunction of the lowest-energy state is still localized within the InAs channel, but with a larger penetration in the InGaAs QW (see Fig. 1(c)): 41% of the probability density is in the InAs channel, 57% is in the InGaAs quantum well and 1.7% is in the InAlAs spacer. Applying a back-gate voltage shifts slightly the QW and the InAs channel in energy relative to the Fermi energy, but does not significantly change the probability density shape (see Fig. 5 in the Appendix). Based on [10] the Rashba coefficient is estimated to be $\alpha = 99.5 \times 10^{-20}$ e m$^2$. The calculations yield an average electric field in the growth direction of $\langle E_z \rangle \approx 4 \times 10^3$ V/cm and a Rashba parameter of $\alpha_\mathrm{R} = \alpha \, \langle E_z \rangle \approx 4 \times 10^{-13}$ eVm.

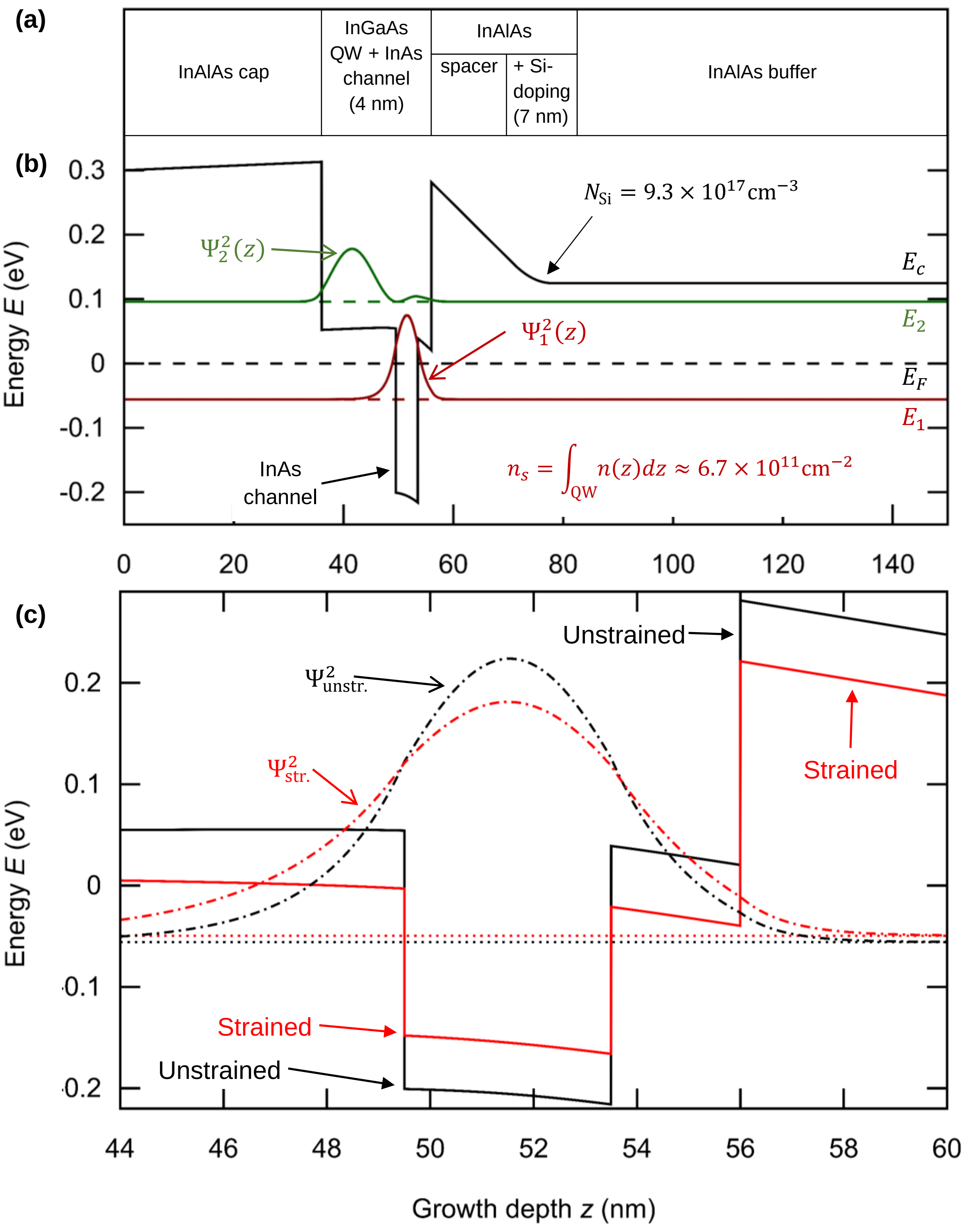

**Fig. 1** Band edges and wavefunction calculation with a 1D Poisson-Schrödinger solver. (a) Layer sequence of the heterostructure. (b) Conduction-band edge $E_C$, energy levels $E_1$ and $E_2$ and probability densities $\Psi_1^2$ and $\Psi_2^2$ of the first two lowest-lying states in the quantum well (QW), and chemical potential $E_F$ (dashed line at $E = 0$), calculated for the heterostructure in (a); $N_{Si}$ is the assumed doping concentration and $n_s$ is the calculated electron sheet density in the QW. (c) Conduction-band edges, energy level and probability density of the lowest states in the QW, for unstrained (black) and strained (red) InAs channel.

### B. Transport in the 2DEG

First, we discuss the magnetotransport measurements of the Hall-bar without constriction which confirm that the QW yields a high-mobility 2DEG. Application of a back-gate voltage changes the carrier density in a range of $10\%$.

The observed negative magnetoresistance at $B < 0.6$ T, see Fig. 2(b), is consistent with the occurrence of weak localization as a quantum correction due to backscattering, which indicates diffusive transport at low fields, and is not affected by the back-gate voltage. The absence of a weak anti-localization in the magnetoresistance at low fields and of a beating pattern in the SdH oscillations at intermediate fields, indicate that the heterostructure and the perpendicular electric field from the back-gate do not induce observable zero-field spin splitting due to spin-orbit-coupling, which confirms previous studies [11]. Zeeman-splitting can be observed for low magnetic fields as $B > 2$ T because of the relatively large $g$-factor.

The existence of topological QHEC transport is evident from the Hall resistance for which the expected QH plateaus occur (see Fig. 2(a)), and the slope of the $R_{xy}(B)$ and the SdH frequencies yield the same sheet carrier density (see Fig. 2(b) and (c)). From the change in $R_{xx}(0)$ and the shift of the SdH minima with varying back-gate voltage $V_{\mathrm{bg}}$, the back-gate voltage changes only the sheet carrier density and not the mobility of $\mu_{\mathrm{H}} \approx 2.4 \times 10^5\ \mathrm{cm}^2/\mathrm{Vs}$ (see Fig. 2(b) and (c)). The change in carrier density with back-gate voltage is consistent with the calculation of the band profiles (see Fig. 1(b)).

The quality of the high-mobility 2DEG is supported by the observation that the magnetoresistance peak of the Zeeman-split LLs at higher magnetic fields (lower energy) is systematically lower than the peak at lower fields (higher energy), see Fig. 2(b) and (c). The lower-field peak corresponds to the chemical potential lying in the higher energy Zeeman-split level, spin-up; the higher-field peak corresponds to spin-down. The reduction of the lower-energy peak is the result of a non-equilibrium population of electrons between the highest-occupied, but partially filled LLs and lowest full LLs along the Hall bar. An equilibrium distribution can be established by inter-edge state coupling only over length scales much larger than the typical Hall-bars. [19, 20]. In Fig. 2(b) and (c), the higher-field peak is significantly reduced for $V_{\mathrm{bg}} = -200$ V, which indicates that the application of a negative back gate voltages minimizes potential fluctuations.

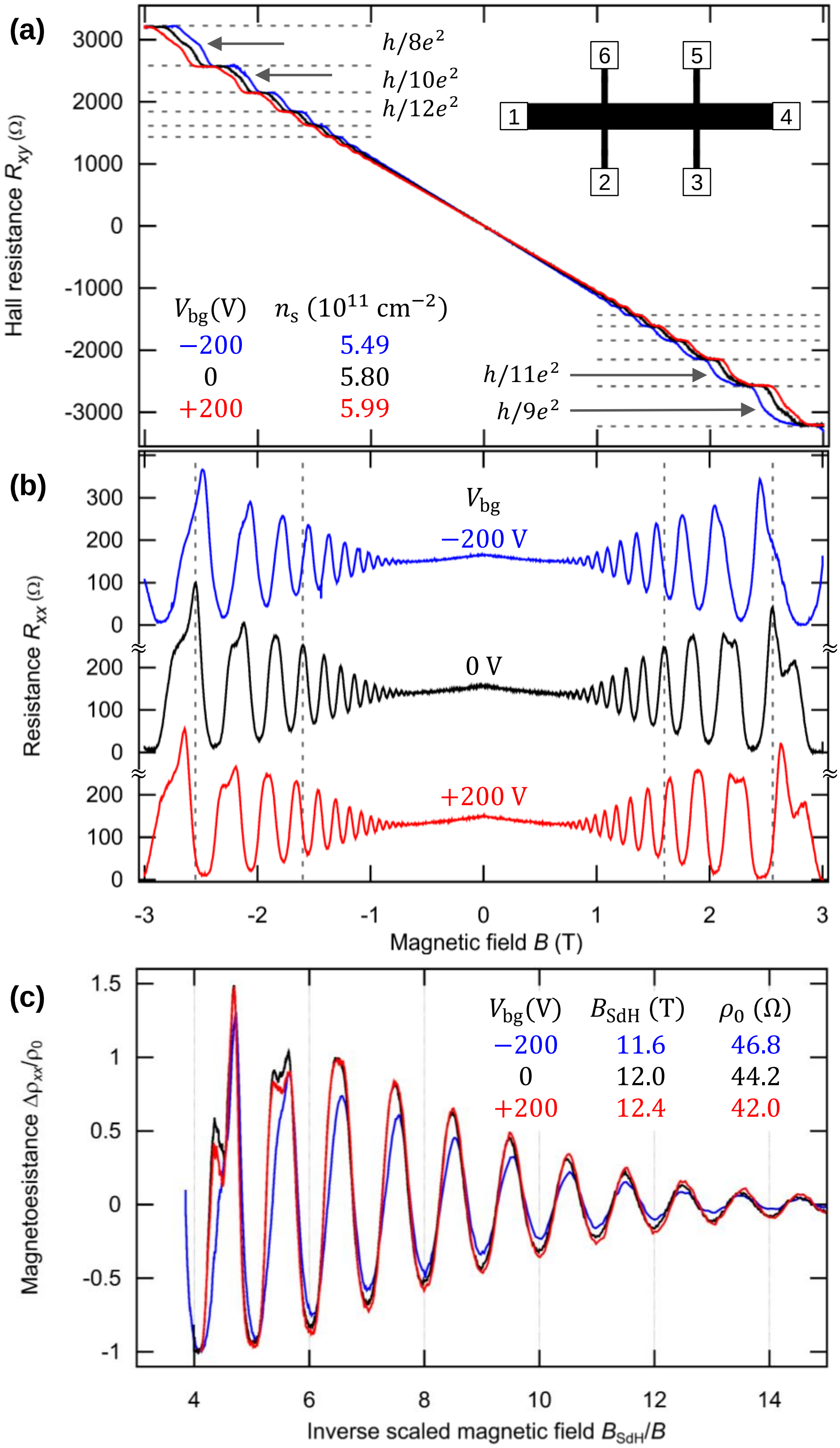

**Fig. 2** Magnetotransport measurements of the Hall-bar at $T_{bath} = 50$ mK and with $I = 2$ nA. (a) Hall resistance $R_{xy}$ as a function of magnetic field $B$, for three back-gate voltages: $V_{bg} = -200, 0, +200$ V. The arrows indicate the Zeeman-split levels. The inset is a schematic depiction of the Hall-bar, with length 65 μm and width 20 μm: $R_{xx} = V_{3,2}/I_{1,4}$ and $R_{xy} = V_{6,2}/I_{1,4}$. (b) Resistance $R_{xx}$ as a function of $B$ for $V_{bg} = -200, 0, +200$ V. (c) Magnetoresistance $\Delta\rho_{xx}/\rho_0$ as a function of the inverse magnetic field scaled by the SdH frequency $B_{SdH}$ of the curve; $\rho_0$ is an average value of the resistivity at $B \approx 0.7$ T.

## C. Transport in the 2DEG with constriction

Forming the constriction in the Hall-bar (see Fig. 3(a)) leaves the positions of SdH oscillations and the edge channel transmission in the QH regime unchanged. The magnetotransport measurements of the Hall-bar with constriction after illumination with an infrared light-emitting diode are shown in Fig. 3(b) and (c) (blue curves). The Hall resistance $R_{xy}$ shows that the Hall-bar with constriction has the same electron density as the Hall-bar without constriction, $n_\mathrm{s} \approx 6.9 \times 10^{11}\ \mathrm{cm}^{-2}$. However, as expected for a constriction with a width $W \approx 4$ µm of the same order as the mean free path $l_\mathrm{e} \approx 2.5$ µm and much larger than the Fermi wavelength $\lambda_\mathrm{F} \approx 30$ nm, we observe an increase of the zero-field resistance and of the SdH peak heights.

Applying a voltage to the in-plane gates in the QH regime allows to switch from an undisturbed transmission of edge channels to a reflection at the constriction, resulting in a filling factor mismatch. The constriction can be electrostatically depleted until a pinch-off occurs at $-15$ V. Cycling the in-plane gate voltages $V_\mathrm{rg} = V_\mathrm{lg}$ between $-15$ V and $+20$ V, while the back-gate is grounded, shows at first a hysteretic behavior, until stability is reached with a reduced carrier density, $n_\mathrm{s} \approx 6.6 \times 10^{11}\ \mathrm{cm}^{-2}$ (see Fig. 3(b) and (c), red curves). In Fig. 3(b), all the QH resistance plateaus show the same quantization at integer fractions of $h/e^2$. The zero-field resistance $R_{xx}(0)$ is determined by the constriction itself: Here, the constriction is quasi-ballistic, because its length $L \approx 4$ µm is of the same order as the mean free path $l_\mathrm{e}$. Hence the resistance increase can be associated to the reduced number of subbands in the constriction and is commonly denoted as Sharvin resistance [21, 22]. From a zero-field resistance increase of $\Delta R_{xx}(0) \approx 90\ \Omega$ we estimate that the number of spin-degenerate subbands in the constriction is $N_\mathrm{c} = \frac{1}{2}\frac{h/e^2}{\Delta R_{xx}(0)} \approx 143$, which yields an effective width $W_\mathrm{eff} = \frac{\pi N_\mathrm{c}}{k_\mathrm{F}} \approx 2.3$ µm with $k_\mathrm{F} \approx 2 \times 10^8\ \mathrm{m}^{-1}$. From this we obtain a depletion length after etching of approximately 0.9 µm. In Fig. 3(c) the SdH oscillations of the Hall-bar with constriction are very similar to the SdH oscillations in the Hall-bar without constriction: However, the peaks are larger due to increased scattering in the constriction. Unlike the SdH oscillations in Fig. 2(b), the peaks corresponding to lower-energy Zeeman-split level are not reduced, because the equilibration length is smaller at higher bath temperatures [23].

As can be seen from Fig. 3(c) the longitudinal resistance $R_{xx}$ is more sensitive to changes due to cycling of the in-plane gate voltages. $R_{xx}(0)$ is higher and yields $\Delta R'_{xx}(0) \approx 390\ \Omega$, $N'_\mathrm{c} \approx 33$ and $W'_\mathrm{eff} \approx 520$ nm, with a doubling of the depletion length. The SdH oscillations are shifted due to the 4% decrease in the electron density in the 2DEG. The decrease is larger in the constriction, because it acts as a barrier potential, increasing the conduction-band edge in the constriction. This results in a filling factor mismatch in the quantum Hall regime and the reflection of one or more edge channels at the constriction [24, 25]. For low magnetic fields the absence of weak localization signals the strong reduction of backscattering in the quasi-ballistic constriction.

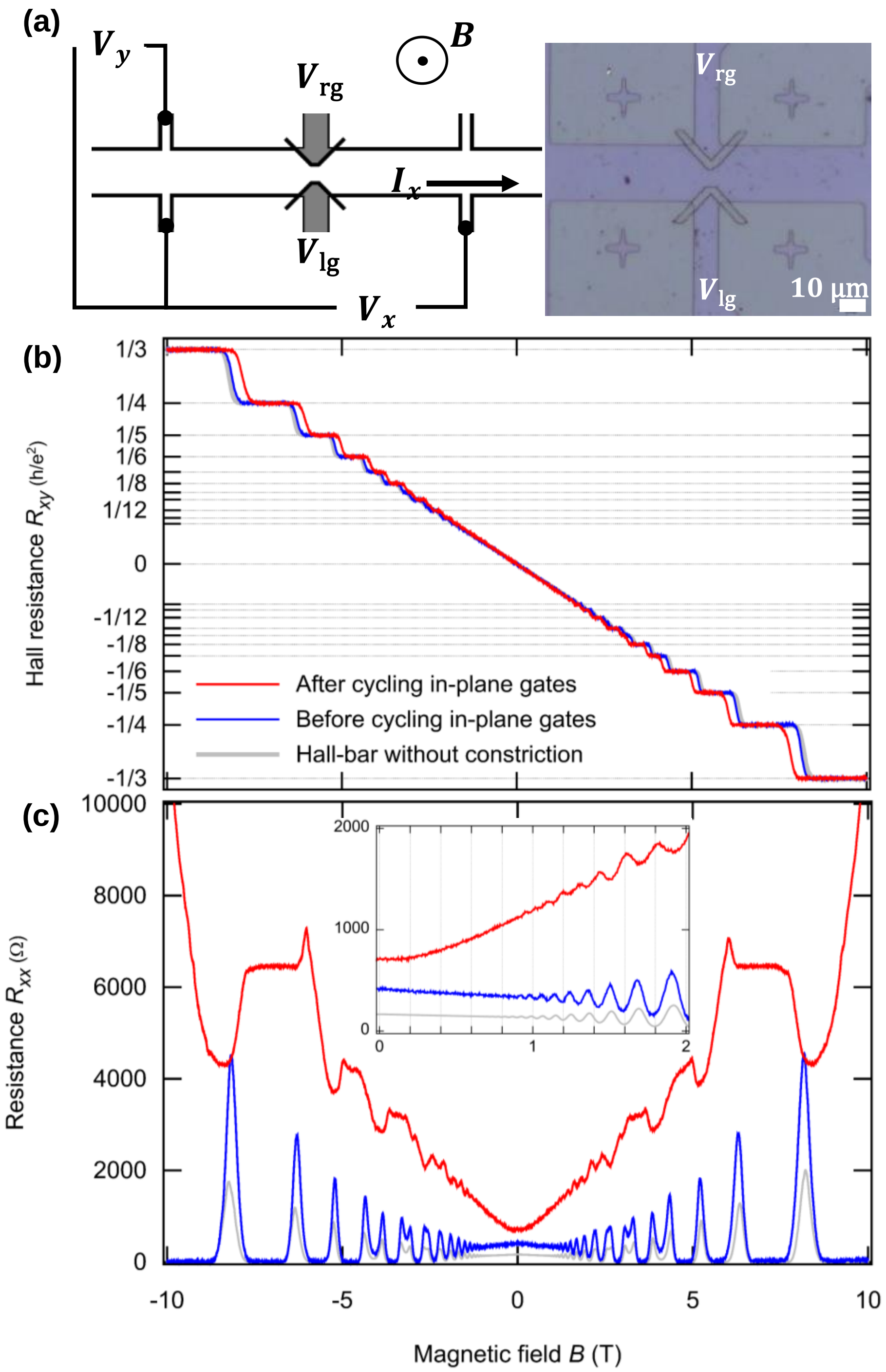

**Fig. 3** Magnetotransport measurements of the Hall-bar with the constriction, after illumination with an infrared LED. The temperature was $T_{\mathrm{bath}} = 250\,\mathrm{mK}$. The measurement current was 5 nA for the constriction and 50 nA for the Hall-bar. All the gates were shorted to ground. (a) Schematic depiction (left) of the measurement setup and optical micrograph (right) of the Hall-bar with constriction, fabricated by wet-chemical etching. The length and the width of Hall-bar are $L_0 = 130\,\mu\mathrm{m}$ and $W_0 = 20\,\mu\mathrm{m}$, respectively. The length and width of the etched constriction are $L \approx 4\,\mu\mathrm{m}$ and $W \approx 4\,\mu\mathrm{m}$, respectively. The two in-plane gates are marked with $V_{\mathrm{rg}}$ and $V_{\mathrm{lg}}$. (b) Hall resistance $R_{xy} = V_y/I_x$ as a function of magnetic field $B$, before (blue) and after (red) cycling the in-plane gate voltages. For comparison, $R_{xy}$ recorded on the Hall-bar without constriction is also depicted (gray). (c) Resistance $R_{xx} = V_x/I_x$ as a function of $B$, before (blue) and after (red) cycling the in-plane gate voltages. $R_{xx}$ for the Hall-bar without constriction is also depicted (gray). The inset shows an enlarged view between $B = 0\,\mathrm{T}$ and $B = 2\,\mathrm{T}$.

### D. Transport in the 2DEG with constriction and transversal in-plane electric field

For the electrostatic control of the transmission and reflection of the QH edge channels, we use *symmetric* in-plane gate voltages $V_{\mathrm{sym}} = V_{\mathrm{rg}} = V_{\mathrm{lg}}$, which change the effective width of the constriction. In Fig. 4(a) the constriction transmits only a few 1D subbands at zero magnetic field for two different scenarios. First, the transmission of edge channels remains largely unchanged for a widened constriction at *positive symmetric* in-plane gate voltages (see $V_{\mathrm{sym}} = 0$ V, grey, and $+10$ V, blue curve), showing only a small change of LL spacing due to a change of electron density. In this case, the depletion of the channel is due to the symmetric in-plane gate voltage and leads to about ten subbands being transmitted at zero-magnetic field. Second, an edge channel mismatch can be induced by narrowing the constriction due to depletion by a *negative symmetric* in-plane gate voltage (see $V_{\mathrm{sym}} = -5$ V, blue dashed curve). In this case, only up to 2 subbands are transmitted in zero-magnetic field. Thereby, it is possible to separate the $\nu = 1$ and $\nu = 2$ edge channels, and the fully spin-polarized current of the $\nu = 1$ edge channel is transmitted through the constriction, at 6 T instead of 10 T. Between $B \approx 7$ T and $B \approx 9$ T all three curves show fluctuations in $R_{xx}$, and corresponding fluctuations can be observed in $R_{xy}$ measured simultaneously. This is consistent with an increased probability of resonant tunneling events between edge states when the chemical potential in the wide 2DEG is shifting from $\nu_0 = 2$ to $\nu_0 = 1$ [26].

To apply a transversal in-plane electric field to the spin-polarized edge channels in the constriction, we use an *asymmetric* in-plane gate voltage, $\Delta V_{\mathrm{as}} = V_{\mathrm{rg}} - V_{\mathrm{lg}}$. The effect is an electrostatic distortion in the constriction, which can be approximated by a saddle-point potential. When $\Delta V_{\mathrm{as}} = 0$ the transversal in-plane electric field is zero and the states are centered on the minimum of the saddle-point potential.

The dependence of the longitudinal resistance on the *transversal* in-plane electric field applied by the asymmetric gate voltage $\Delta V_{\mathrm{as}}$ is observed in the transition between Zeeman-split LLs (QH plateaus), as can be seen in Fig. 4(b) for the case with a filling factor mismatch for the transition from $\nu = 1$ to $\nu = 2$, and in Fig 4(c) for the case without mismatch between the wide Hall bar and the constriction in the transition from $\nu = 1$ to $\nu = 2$ and from $\nu = 2$ to $\nu = 3$. The following observations on the orientation dependence of the transversal in-plane electric field on the longitudinal resistance do not find an explanation in a purely electrostatic response of charges.

For the case of the transmission without a mismatch the shift of the longitudinal resistance features most prominently occurs at the magnetic field of about 5 T, which is marked by (ii) in Fig. 4(c). This shows that the transversal in-plane electric field affects magnitude and the magnetic field position of the resistance increase depending on the spin state. When the chemical potential is in the spin-up Zeemann-split LL (ii), the backscattering is increased (green curve) or suppressed (red curve) depending on the transversal in-plane field orientation. This indicates that the transversal in-plane electric field changes the density of states of the spin-split LLs.

Together with the peak position for the spin-down Zeemann split LL, marked by (i) in Fig. 4(c), the magnetic field shift can be measured to be $\Delta B \approx 0.1$ T, which is consistent with an effective magnetic field of $B_{\mathrm{eff}} = (v_{\mathrm{F}} E_{\mathrm{transv}})/c^2 \approx (6 \times 10^7\ \mathrm{cm/s} \cdot 2 \times 10^4\ \mathrm{V/cm})/c^2 = 0.13$ T for $\Delta V_{\mathrm{as}} = 20$ V, normal to the 2DEG plane, where $v_{\mathrm{F}}$ is the Fermi velocity and $E_{\mathrm{transv}}$ is the transversal in-plane electric field. From this we determine the in-plane spin-orbit coupling parameter $\alpha_{\mathrm{so}}(\Delta V_{\mathrm{as}} = 20\ \mathrm{V}) = g^* \mu_{\mathrm{B}} \Delta B / 2k_{\mathrm{F}} \approx 2 \times 10^{-13}$ eVm and the corresponding in-plane coefficient $\alpha_{\text{in-plane}} \approx 1 \times 10^{-19}$ e m$^2$ for the Hall bar with etched constriction. This value is smaller by a factor 10 compared to the calculations for this heterostructure in Section 2A. The difference can partly be accounted for by the wavefunction penetration into InGaAs, possibly by the strain in InAs and perhaps by the etched constriction which were all not included in the calculations.

A similar resistance increase and decrease depending on the orientation of the transversal in-plane field can be seen for the transition from $\nu = 2$ to $\nu = 1$ at 8 T marked by (iii) and (iv) in Fig. 4(c). Here, increased backscattering sets in at lower magnetic fields (green curve), while the backscattering is strongly suppressed (red curve) at the resistance maximum. Because in the transition to $\nu = 1$ only one spin state remains available for the final state after scattering processes, the effect of the orientation of the transversal in-plane electric field is not symmetric with respect to the two spin states. This situation is similarly in the $\nu = 2$ to $\nu = 1$ transition for the case of a mismatch, see Fig. 4(b), marked with (ii).

From the above considerations, we conclude that, although spin-orbit coupling does not lead to observable spin-splitting of subbands at zero-magnetic field as discussed in Section 2B, it needs to be taken into account when transversal in-plane fields are applied in the QH regime for 2DEGs which are subject to lateral confinement in the order of the mean free path and below. Due to spin-orbit coupling the orientation of transversal in-plane electric fields with respect to the current flow can affect the spin-dependent scattering between QHECs situated in the wide Hall bar and the constriction. This becomes most evident when increased backscattering is possible as in in the transition between the LLs (QH plateaus).

These experiments show that the transport in QH regime under application of micro- or nanopatterned gates and/or constrictions requires the thorough investigation of the undisturbed 2DEG (wide Hall-bars) as a reference system. Furthermore, in particular for 2DEGs with spin-orbit interaction, such as InAs-based heterostructures, the application of in-plane gates may serve as a powerful tool to explore spin-dependent phenomena by transversal electric fields.

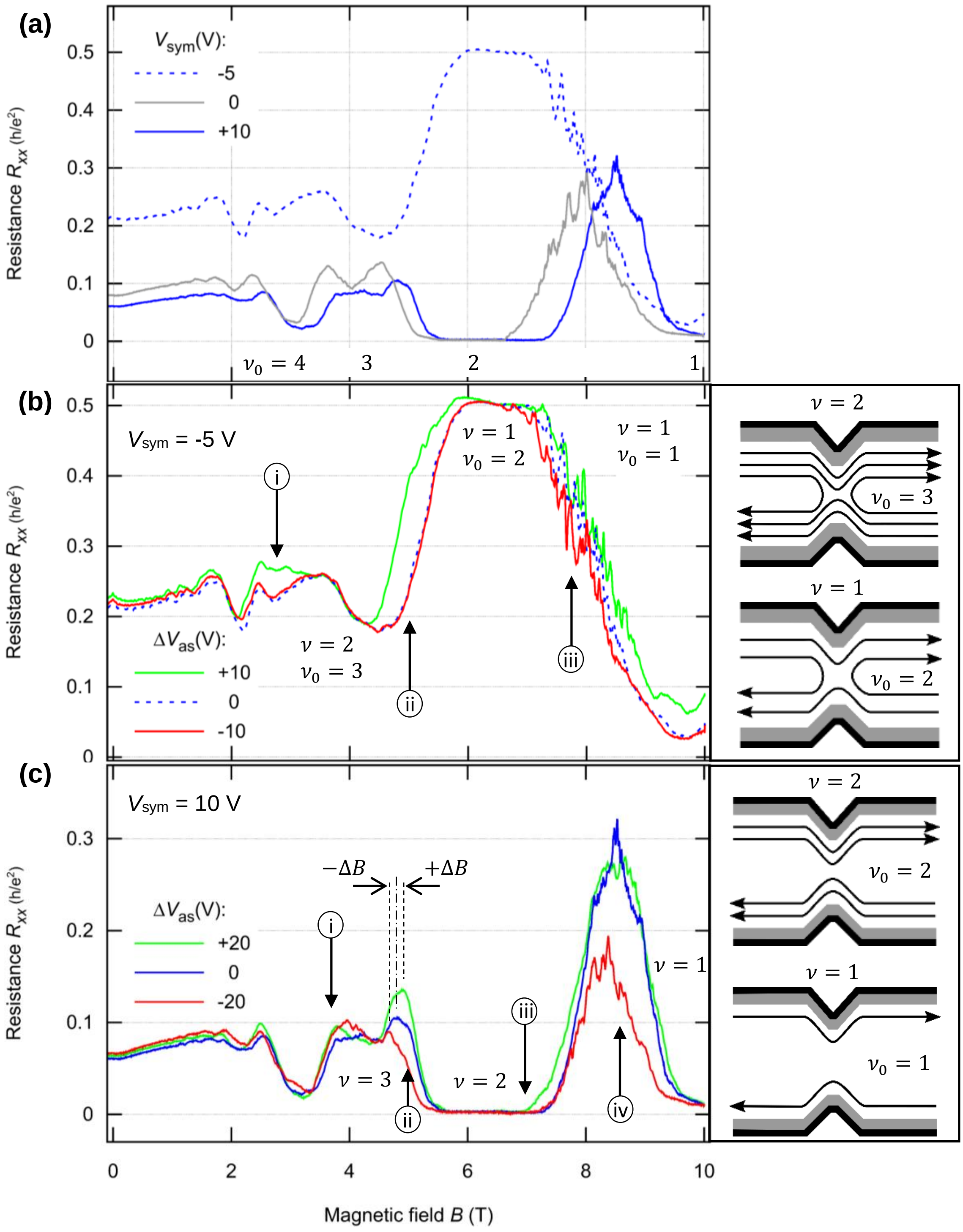

**Fig. 4** Magnetotransport measurements of the Hall-bar with the etched constriction, after further illumination with an infrared LED and cycling of the in-plane gates. The temperature was $T_{bath} = 250\ \mathrm{mK}$ and the current $I = 5\ \mathrm{nA}$. (a) Resistance $R_{xx}$ as a function of magnetic field $B$ for symmetric in-plane gate voltages $V_{rg} = V_{lg} = V_{sym}$. The filling factor of the wide 2DEG connected to the constriction is indicated by $\nu_0$. (b) and (c) $R_{xx}$ as a function of $B$ for asymmetric in-plane gate voltages: $V_{rg} = V_{sym} + (\Delta V_{as}/2)$ and $V_{rg} = V_{sym} - (\Delta V_{as}/2)$. $\Delta V_{as}$ is proportional to the transversal in-plane electric field in the constriction. The filling factor in the constriction is indicated by $\nu$. The sketches on the right depict the constriction and the edge channels for various cases: with mismatch (b) and without mismatch (c). The arrows marked with *i*, *ii*, *iii* and *iv* are discussed in the text. The dash-dotted line indicates a peak in the blue curve, from which the shifts $+\Delta B$ (green) and $-\Delta B$ (red) of peaks marked with dashed lines are measured.

# III. Conclusions

For a high-mobility 2DEG with a large Landé $g$-factor of 16 in an InAs-based quantum well it was demonstrated that for an etched micro-constriction with in-plane gates the QH regime with only one spin-polarized QHEC (filling factor $\nu = 1$) can be reached at a moderate magnetic field of 9 T (no filling factor mismatch) or lower, at 6 T (for filling factor mismatch). The electrostatic control by symmetric voltages applied to the in-plane gates allows to tune the transmission and reflection of QH edge channels. The orientation of a transverse in-plane electric field by asymmetric voltages applied to the in-plane gates affects the magnetic field position and the magnitude of the longitudinal resistivity. This strongly indicates that the spin-dependent backscattering between spin-polarized QH edge channels in the transition between Landau levels (QH plateaus) is affected by spin-orbit coupling.

## Acknowledgments

We thank Christian Riha for scientific and technical support. OC, JB, and SFF acknowledge support from the Deutsche Forschungsgemeinschaft under grant INST 276/709-1.

## Author Declarations

### Conflict of Interest

The authors have no conflicts to disclose.

### Author Contributions

**Olivio Chiatti**: Conceptualization (equal); Data curation (equal); Formal analysis (equal); Investigation (equal); Methodology (equal); Project Administration (equal); Resources (equal); Validation (equal); Visualization (equal); Writing – Original Draft Preparation (equal); Writing – Review & Editing (equal). **Johannes Boy**: Data curation (equal); Formal analysis (equal); Investigation (equal); Writing – Review & Editing (supporting). **Christian Heyn**: Resources (equal); Writing – Review & Editing (supporting). **Wolfgang Hansen**: Resources (equal); Methodology (supporting); Writing – Review & Editing (supporting). **Saskia F. Fischer**: Conceptualization (equal); Funding acquisition (lead); Methodology (equal); Project Administration (equal); Supervision (lead); Validation (equal); Visualization (equal); Writing – Original Draft Preparation (equal); Writing – Review & Editing (equal).

## Data Availability Statement

The data that support the findings of this study are available from the corresponding author upon reasonable request.

# Appendix

Table I: Layer sequence and parameters used for the calculations by the 1D Poisson-Schrödinger solver: growth depth $z$ from the surface, thickness $\Delta z$ of each layer, material of the layer, bandgap $E_\mathrm{g}$, conduction band offset $\Delta E_\mathrm{c}$ relative to GaAs, electron effective mass $m_\mathrm{e}^*$, Schottky barrier height $\Delta E_\mathrm{S}$, fraction of the probability density $\Psi_1^2$ of the lowest-energy electronic wavefunction in the QW. In bold are values for strained InAs.

| Growth depth $z$ (nm) | Layer thickness $\Delta z$ (nm) | Material | Lattice parameter $a$ (nm) | Bandgap $E_g$ (eV) | Conduction band offset $\Delta E_c$ (eV) | Electron effective mass $m_e^*$ ($m_e$) | Schottky barrier height $\Delta E_S$ (eV) | Fraction of $\Psi_1^2$ |
|---|---|---|---|---|---|---|---|---|
| 0.0 | 36.0 | $In_{0.75}Al_{0.25}As$ | 0.5951 | 0.956 | -0.406 | 0.0478 | 0.30 | 0.0016 **0.0046** |
| 36.0 | 13.5 | $In_{0.75}Ga_{0.25}As$ | 0.5948 | 0.603 | -0.667 | 0.0346 | 0.05 | 0.3065 **0.3758** |
| 49.5 | 4.0 | InAs: unstr. **strained** | 0.6050 **0.5950** | 0.417 **0.450** | -0.922 **-0.812** | 0.0260 **0.0260** | 0.00 **0.00** | 0.4611 **0.3862** |
| 53.5 | 2.5 | $In_{0.75}Ga_{0.25}As$ | 0.5948 | 0.603 | -0.667 | 0.0346 | 0.05 | 0.1672 **0.1584** |
| 56.0 | 15.0 | $In_{0.75}Al_{0.25}As$ | | | | | | |
| 71.0 | 7.0 | $In_{0.75}Al_{0.25}As$: Si | 0.5951 | 0.956 | -0.406 | 0.0478 | 0.30 | 0.0623 **0.0743** |
| 78.0 | 430 | $In_{0.75}Al_{0.25}As$ | | | | | | |

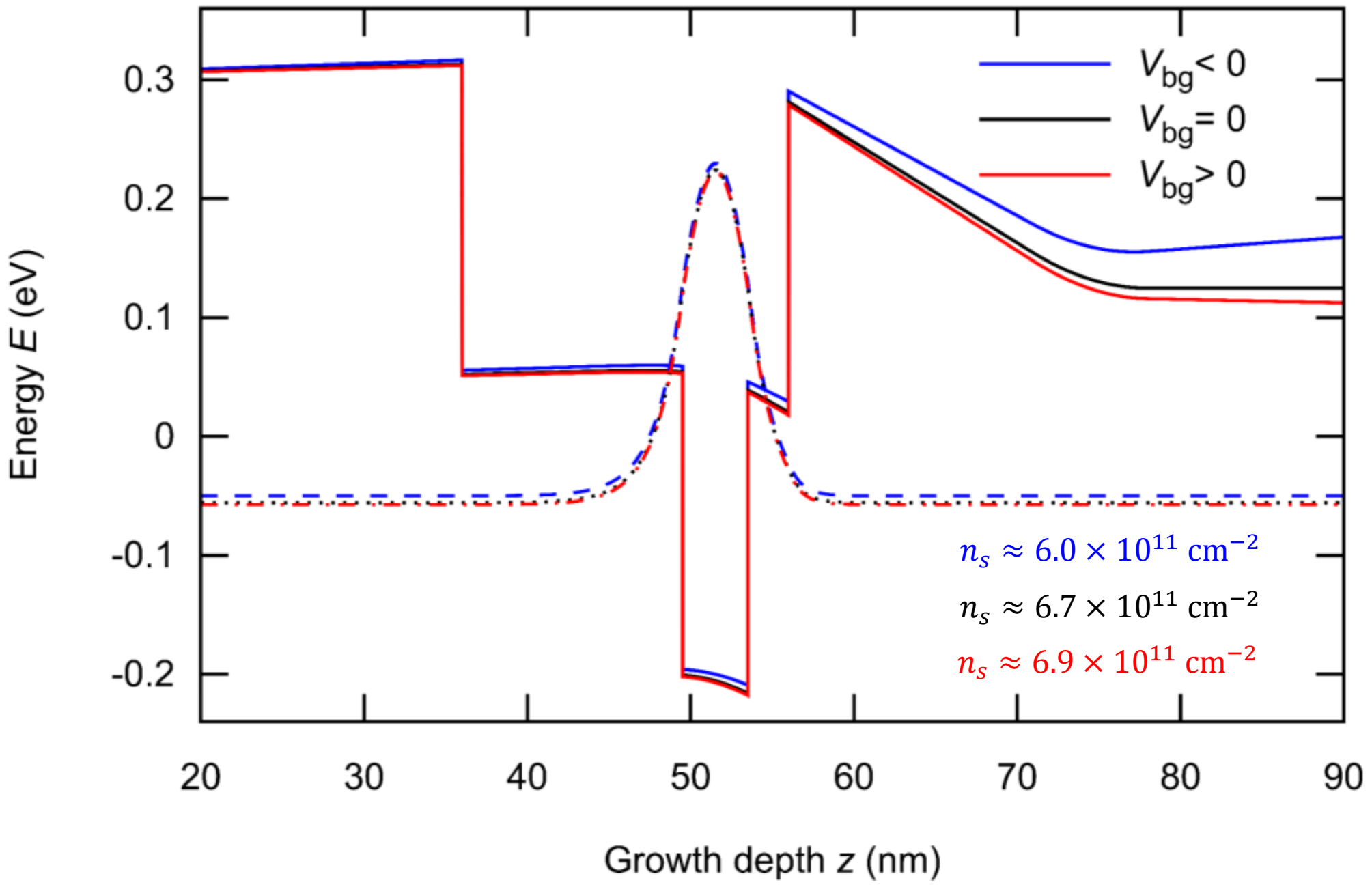

**Fig. 5** Conduction band edge and lowest-energy electronic wavefunction for three values of the back-gate voltage $V_\mathrm{bg}$. The back-gate voltage was modelled by fixing the chemical potential at a depth of approximately 1 µm. The sheet density $n_\mathrm{s}$ varies with back-gate voltage in the same range as in the experiments.